\documentclass[12pt]{article}
\usepackage{graphicx}
\usepackage{amssymb} 

\newcommand\pubnumber{CIPANP2018-Deaconu}
\newcommand\pubdate{\today}

\def\kicp{Kavli Institute for Cosmological Physics, Dept. of Physics, Enrico
Fermi Institute, University of Chicago, Chicago, IL 60637, USA }

\def\Title#1{\begin{center} {\Large #1 } \end{center}}
\def\Author#1{\begin{center}{ \sc #1} \end{center}}
\def\Address#1{\begin{center}{ \it #1} \end{center}}

\newcommand\pubblock{\rightline{\begin{tabular}{l} \pubnumber\\
  \pubdate \end{tabular}}}
\newenvironment{Abstract}{\begin{quotation} }{\end{quotation}}
\newenvironment{Presented}{\begin{quotation} \begin{center} 
  PRESENTED AT\end{center}\bigskip 
 \begin{center}\begin{large}}{\end{large}\end{center} \end{quotation}}
\def\Acknowledgements{\bigskip \bigskip \begin{center} \begin{large}
  \bf ACKNOWLEDGEMENTS \end{large}\end{center}}

\def\beq{\begin{equation}}
\def\eeq#1{\label{#1}\end{equation}}
\def\eeqn{\end{equation}}

\def\beqa{\begin{eqnarray}}
\def\eeqa#1{\label{#1}\end{eqnarray}}
\def\eeqan{\end{eqnarray}}

\let\bar=\overbar

\def\Dslash{\not{\hbox{\kern-4pt $D$}}}
\def\dslash{\not{\hbox{\kern-2pt $\del$}}}

\def\msb{{\bar{\ssstyle M \kern -1pt S}}}

\begin{document}
\begin{titlepage}
\pubblock

\vfill
\Title{Recent Results from ANITA}
\vfill
\Author{Cosmin Deaconu for the ANITA Collaboration}
\Address{\kicp}
\vfill
\begin{Abstract}

The ANtarctic Impulsive Transient Antenna (ANITA) long-duration balloon payload
searches for Askaryan radio emission from ultra-high-energy ($>10^{18}$ eV)
neutrinos interacting in Antarctic ice. ANITA is also sensitive to geomagnetic
radio emission from extensive air showers (EAS). This talk summarizes recently
released results from the third flight of ANITA, which flew during the
2014-2015 Austral summer. The most sensitive search from ANITA-III identified
one neutrino candidate with an \textit{a priori} background estimate of
0.7$^{+0.5}_{-0.3}$. When combined with previous flights, ANITA sets the best
limits on diffuse neutrino flux at energies above $\sim10^{19.5} eV$.
Additionally, ANITA-III searches identified nearly 30 EAS candidates. One
unusual event appears to correspond to an upward-going air shower, similar to
an event from ANITA-I. 

\end{Abstract}
\vfill
\begin{Presented}
Thirteenth Conference on the Intersections of Particle and Nuclear Physics \\
May 29-June 3, 2018
\end{Presented}
\vfill
\end{titlepage}
\def\thefootnote{\fnsymbol{footnote}}
\setcounter{footnote}{0}

\section{Introduction}

Interactions of ultra-high-energy cosmic rays with the cosmic microwave
background are expected to produce a population of EeV-scale ``cosmogenic"
neutrinos\cite{Beresinsky:1969qj}. The ANtarctic Impulsive Transient Antenna
(ANITA) long-duration balloon experiment seeks to detect these neutrinos by
monitoring Antarctic ice for the wideband (180-1200 MHz) impulsive Askaryan
emission produced by neutrino-induced-cascades\cite{Askaryan:1962hbi}. Glacial
ice is abundant in Antarctica and has excellent radio propagation properties.
At these energies, the Earth is opaque to neutrinos in the Standard Model, so most
neutrinos can only skim the earth. Due to the event geometry, the radial
polarization of Askaryan emission, and preferential vertically-polarized
transmission at the ice-air interface, radio emission from neutrinos is
expected to be primarily vertically-polarized.

ANITA is also sensitive to impulsive wideband emission from extensive air
showers (EAS). In this case, the majority of the emission is from the Earth's
geomagnetic field, which is approximately vertical in Antarctica, resulting in
a primarily horizontally-polarized signal. Due to ANITA's altitude, the primary
EAS development is below the payload, so, except for very grazing cosmic-rays
that miss the ice entirely, the emission from cosmic-ray induced EASs will be
reflected off the ice, introducing a polarity flip~\cite{Hoover:2010qt}.

\section{The ANITA-III flight}

\begin{figure} 
 \includegraphics[width=4.2in]{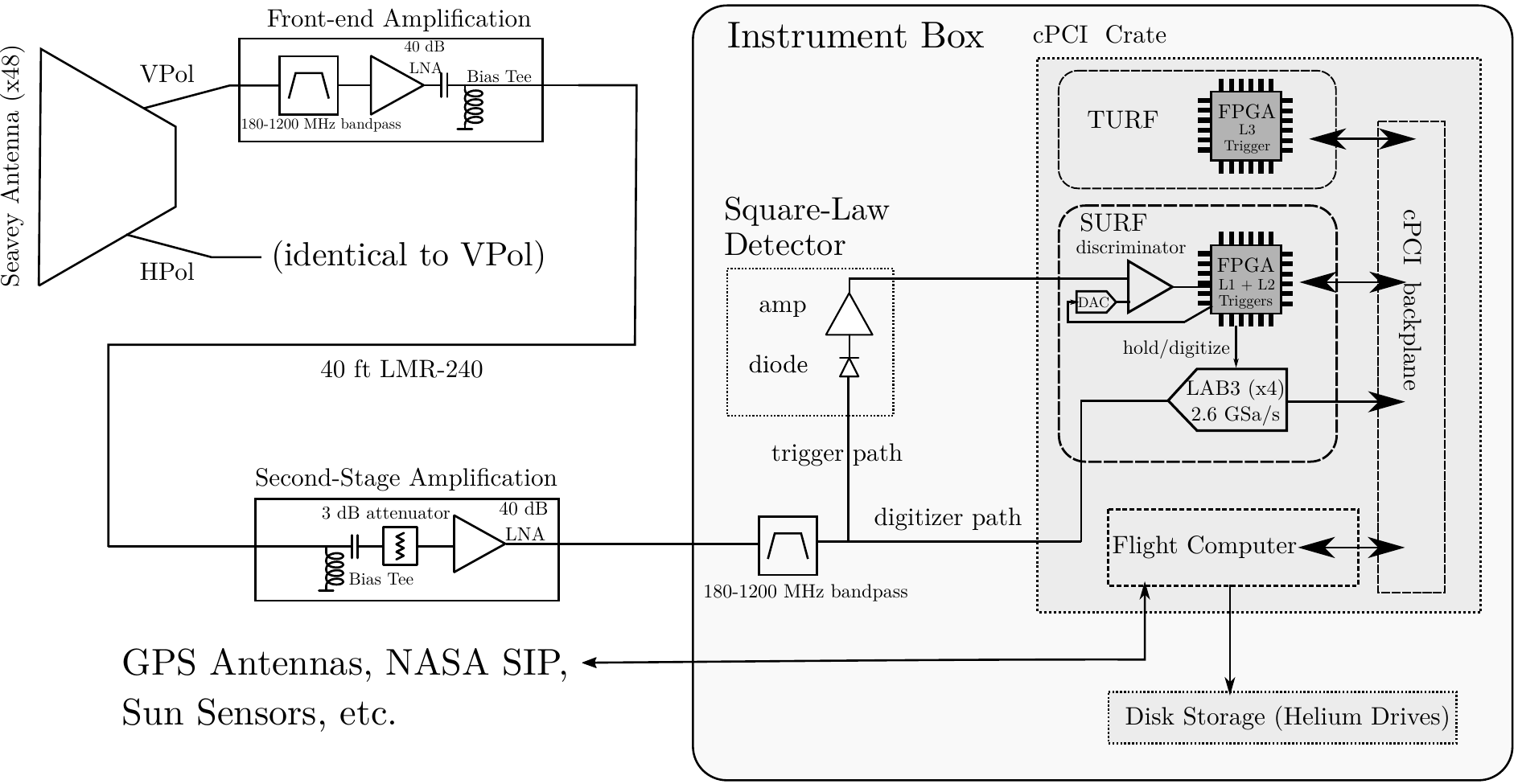}
 \vline 
 \hspace{0.05in}
 \includegraphics[width=1.6in]{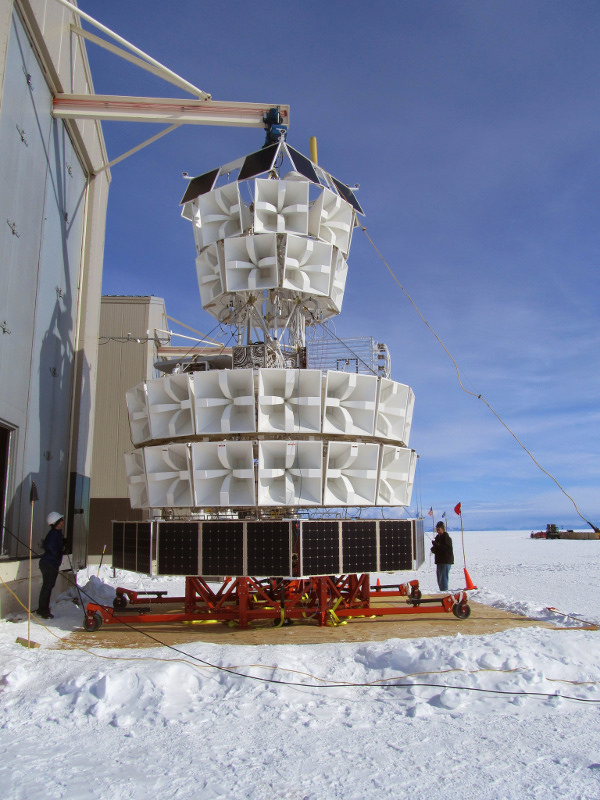}
 \caption{ Left: A block diagram of the ANITA-III instrument. Signals from each antenna are amplified, and then split into a trigger and digitizer path. Right: ANITA-III prior to launch at the Long-Duration Ballooning Facility in Antarctica.} 
 \label{fig:system}
\end{figure}

The third ANITA flight (ANITA-III) began on December 18, 2014 and flew for 22
days before termination. ANITA-III has a similar design to previous flights, but
has additional antennas and a different trigger. ANITA-III has 48
dual-polarization wideband horn antennas, arranged in an azimuthal pattern such
that that there are 16 azimuthal sectors, with a top, middle and bottom antenna
each. ANITA-III is depicted both schematically and photographically in Fig.~\ref{fig:system}. 

The first-level per-channel trigger compares the output of a tunnel diode
square-law detector with a threshold that is dynamically adjusted to keep the
first-level trigger rates at 450 kHz. The second-level level trigger requires a
coincidence within at least two antennas within a sector to trigger. The
coincidence window is set for each antenna pair and enforces causality. The
global trigger requires a coincidence of two adjacent azimuthal sectors.
ANITA-III triggered on horizontal and vertical polarization independently. The
global rate was typically around 50 Hz.

When a global trigger is issued, roughly 100 ns of data is recorded in all
channels using a LAB3 switched-capacitor array digitizer~\cite{Varner:2007zz}.
To prevent noisy anthropogenic sources from dominating the trigger, sectors are
masked from trigger if the global rate from the sector is too high. Due to
continuous-wave (CW) anthropogenic signals from satellites, the
northern-facing half of the payload was masked for much of the flight in ANITA-III.

\section{Analysis Methods} 

Several independent searches for impulsive signals from EASs or
neutrinos were performed on the data from ANITA-III~\cite{a3nu}~\cite{a3up}.
ANITA-III recorded 70 million triggers, the overwhelming majority of which are
thermal noise, anthropogenic signals, or ``payload blasts,"---RFI from the
payload itself. Much of the data is contaminated by CW, even when the CW is not
responsible for the trigger. 

Each search made different choices, but many of the steps are similar and are
summarized here. First, waveforms are filtered to remove CW contamination that
can confuse the analysis. Then, an interferometric map is created, where each
incoming plane wave direction is probed and checked for consistency with
inter-antenna timings. The best directions are chosen and a coherent waveform
is formed in that direction from all antennas pointing nearby.  From the map
and coherent waveform, a number of observables are generated which are combined
to distinguish impulsive signals from thermal noise (which is incoherent) and
payload blasts (which are not plane-wave like). For training, Monte Carlo
neutrinos and/or calibration pulsers are used to as a signal sample and a
sideband region is used for a background sample.  Passing events are traced
back to the ice and events are checked for isolation from other events as it us
unlikely that multiple physics events will come from the same place. 

For a given set of cuts, Monte Carlo and events from calibration pulsers are
used to estimate the analysis efficiency. The background is estimated from the
data, either by using distributions of observables or a sideband. 

Three searches were optimized for the Askaryan neutrino vertically-polarized
channel, but also searched for EAS candidates in the horizontally-polarized
channel. A dedicated EAS search additionally checks waveforms against a signal
shape template validated using EAS signals observed in previous flights, which
further lowers the expected background.

\section{Results}

 \begin{figure}
 \includegraphics[width=3.9in]{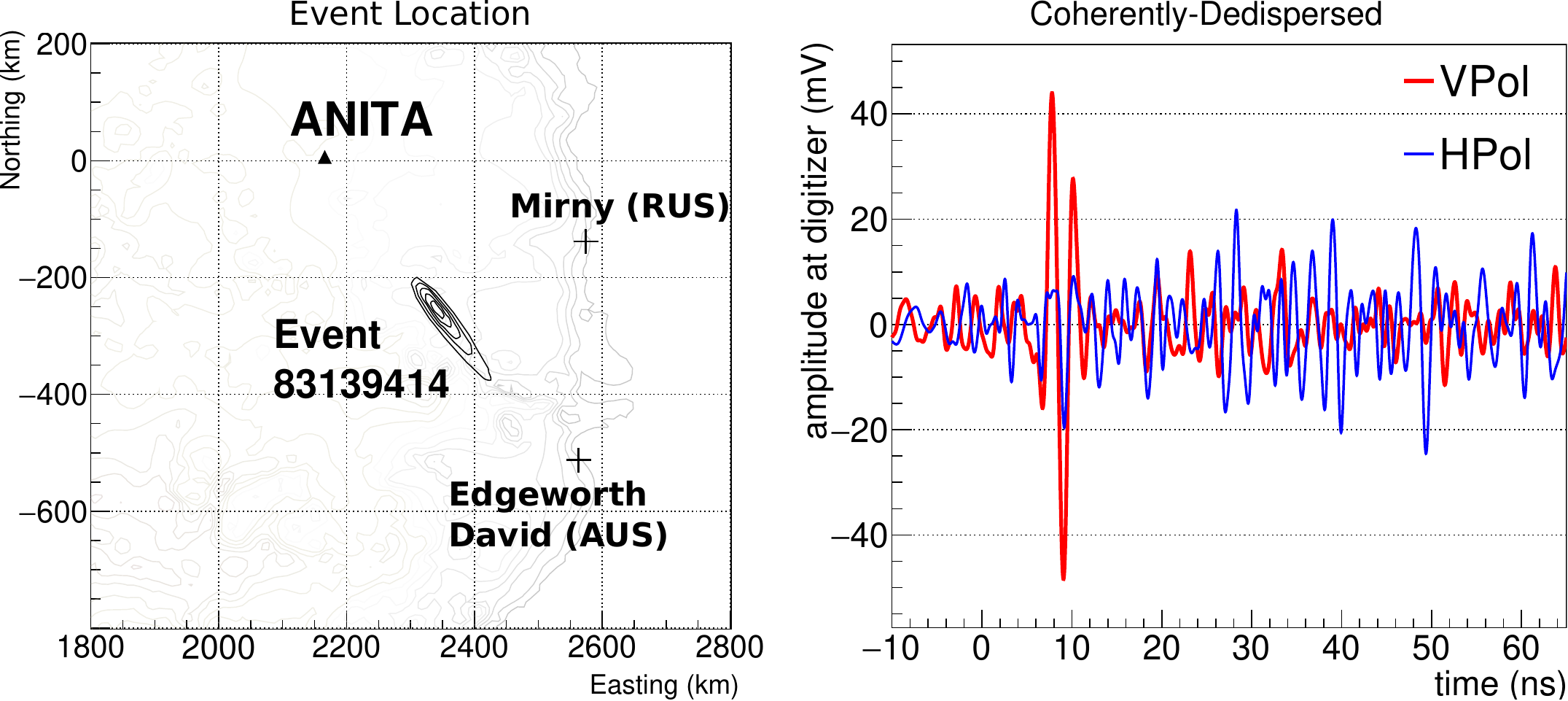} 
 \includegraphics[width=2in]{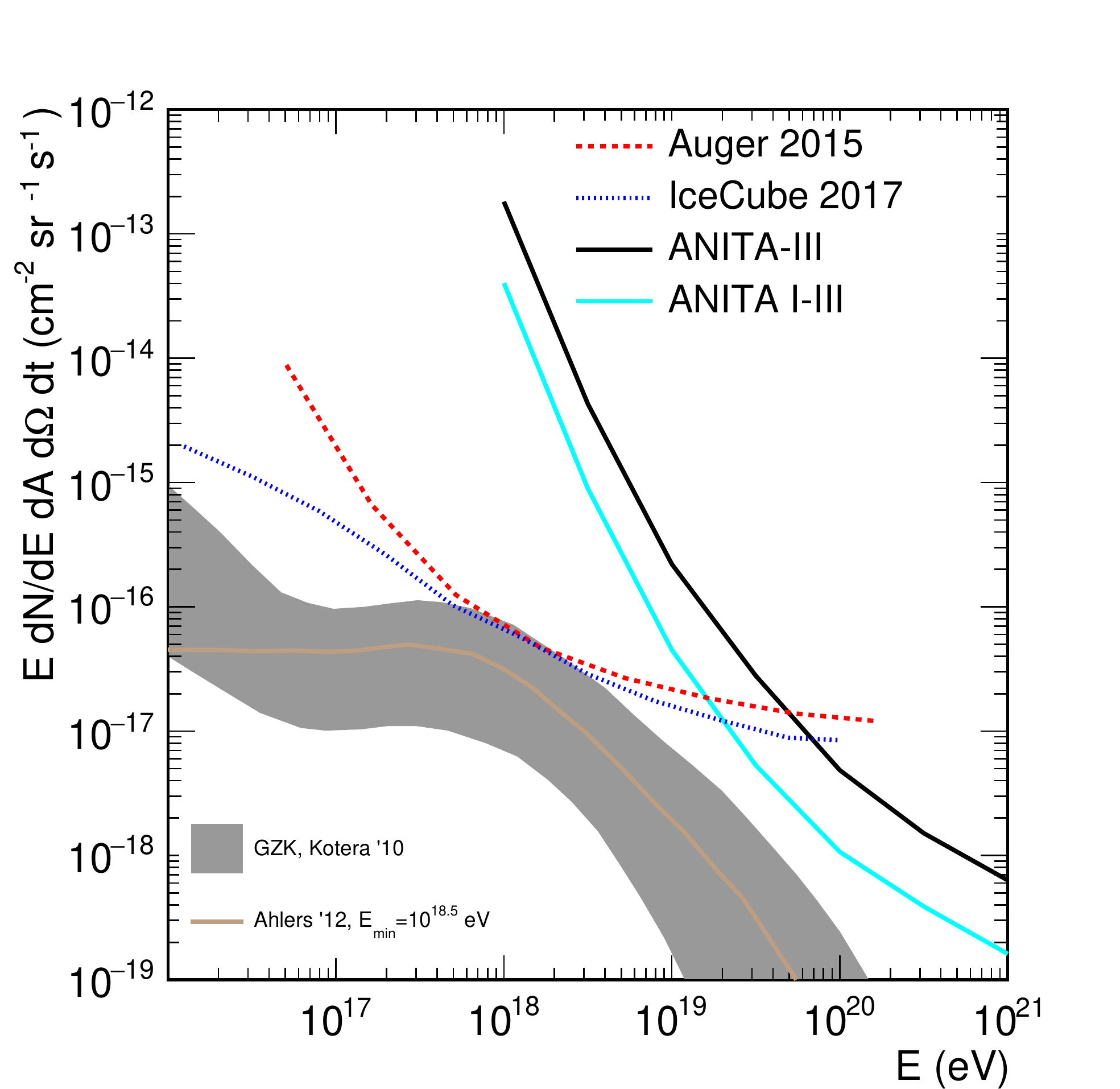}
 \caption{Left: The event localization of isolated, impulsive, vertically-polarized event 83139414. Center: The dedispersed, coherently-summed waveform for event 83139414. Right: The diffuse flux upper limit from ANITA III alone and ANITA I-III combined, interpreting the candidate as a background.} 
 \label{fig:candidate_event} 
 \end{figure}

 \begin{figure} 
 \includegraphics[width=2.5in]{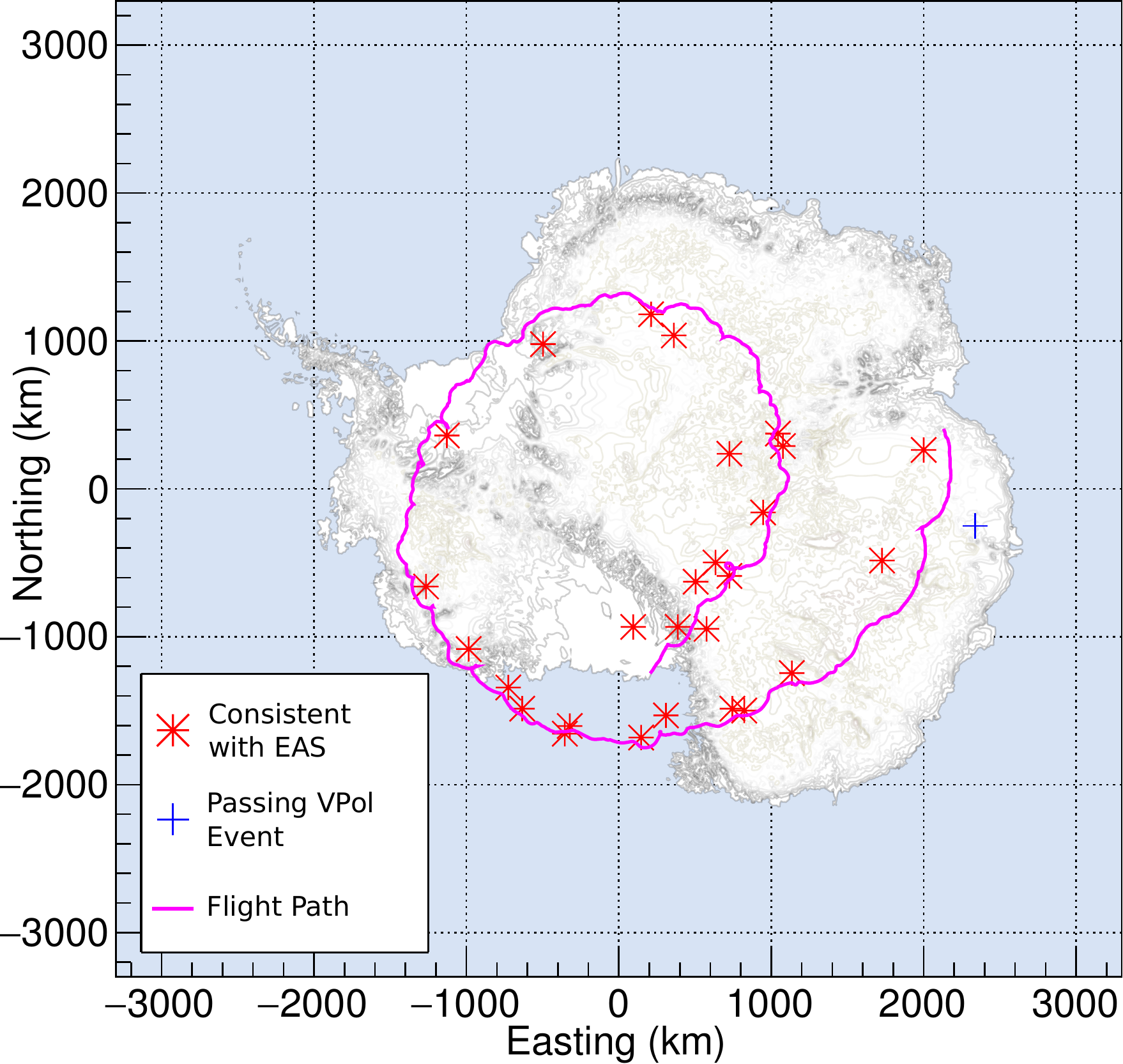} 
 \includegraphics[width=3in]{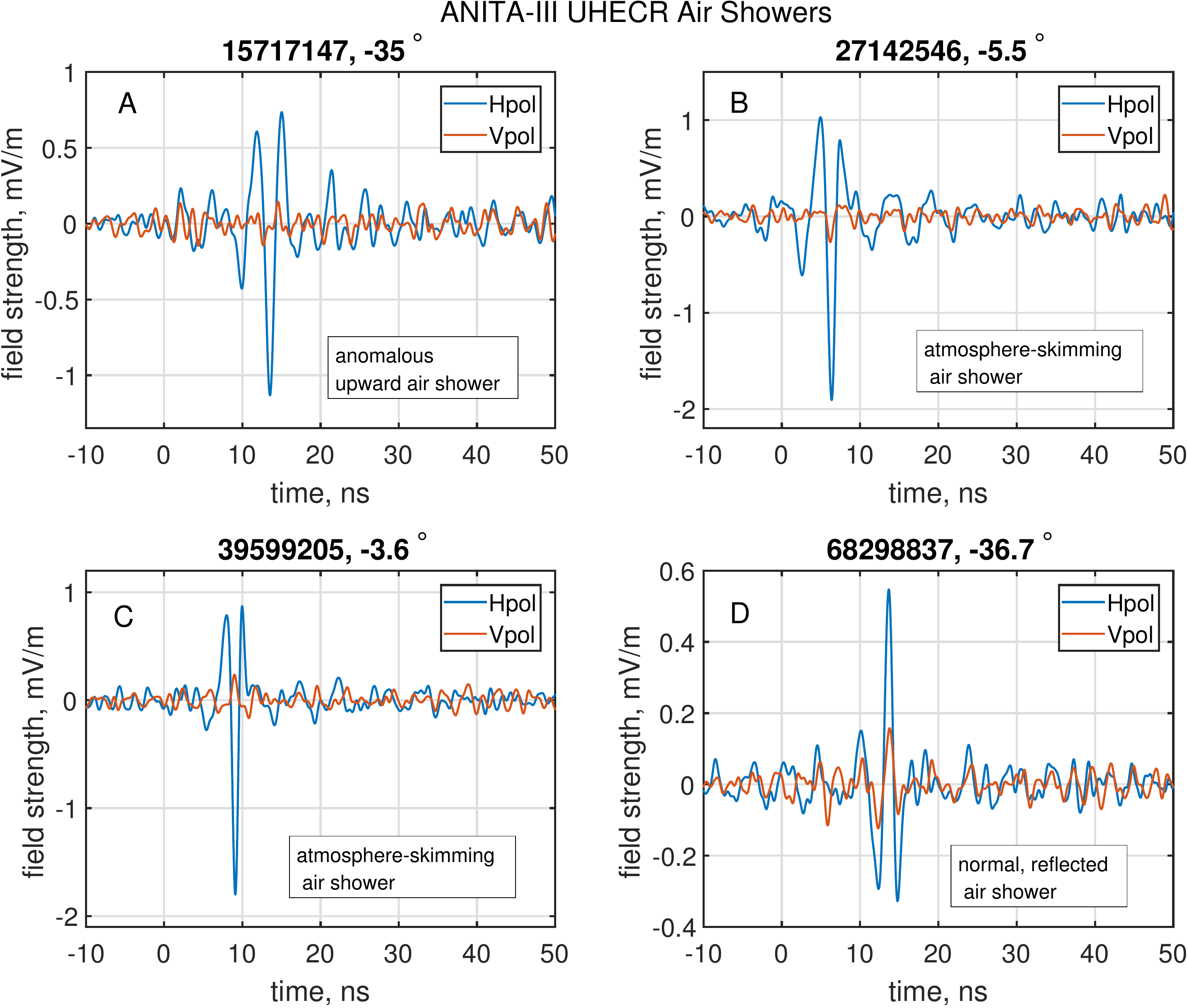} 
 \caption{
 Left: The locations of all events identified by the various searches. Right:
 The coherently-summed, dedispersed signal from the anomalous apparently
 upward-going shower compared to two atmosphere-skimming showers and one
 reflected shower. } \label{fig:map_events} 
 \end{figure}

In the vertically-polarized channel, the Askaryan neutrino with the best
expected pre-unblinding sensitivity yielded one signal candidate (event
83139414) on an estimated background of $0.7^{+0.5}_{-0.3}$ events. While
consistent with the background estimate, the signal candidate
(Fig.~\ref{fig:candidate_event}, left two panels) is in deep ice, very isolated
from other events and known sources of human activity, and has a signal shape
and polarization consistent with expectations from simulation. Interpreting the
event as a background, we set a limit depicted on the right side of
Fig.~\ref{fig:candidate_event}. ANITA I-III combined set the best limit on
diffuse cosmogenic neutrino flux above $\sim10^{19.5} eV$. 

Between all analyses, 28 isolated impulsive horizontally-polarized events
were found that are consistent in pulse shape and polarization with an EAS
origin. The candidate locations are mapped on the left side of
Fig.~\ref{fig:map_events}. Two events were above the horizon and 26 events
were below. Of those, event 15717147 is notable because it has
polarity consistent with an upward shower but comes from 35 degrees below the
horizontal, very clearly below the horizon. The background estimate from the
dedicated EAS search for this event is $\lesssim 0.01$ events, based on the
characteristics of nearby triggers. Together with a similar previous event from
ANITA-I~\cite{upward}, this event topology defies a straightforward explanation. A
decaying $\tau$ from a $\nu_{\tau}$ interacting in the ice could produce an
upward shower, but a SM $\nu_{\tau}$ would have been unlikely to make it
through so much of the Earth to produce such an energetic shower. Moreover,
such an observation from a diffuse flux would be in tension with results from
other experiments~\cite{tension}.

\section{Conclusion}

The ANITA experiment has set world-leading limits on diffuse high-energy neutrino
flux. Moreover, two anomalous events have been found that require more detailed
understanding. The fourth flight of ANITA flew in December 2016,
with upgrades to mitigate CW noise from satellites. Analysis of ANITA-IV data is
ongoing. Planned upgrades for future ANITA flights include improved digitizers and a lower-threshold interferometric trigger.

\pagebreak
\Acknowledgements

This author acknowledges support from NASA, the NSF, the University of Chicago Research Computing Center, and the Kavli Institute for Cosmological Physics at the University of Chicago.

\end{document}